\documentclass[preprint]{ptephy_arxiv}

\preprintnumber{XXXX-XXXX} 
\usepackage{siunitx}
\usepackage{ulem}
\usepackage{comment}
\usepackage{url}

\newcommand{\ee}{$e^+e^- $} 
\newcommand{\phiee}{$\phi \rightarrow e^+e^- $}

\newcommand{\MI}[1]{{\color{blue}#1}}

\renewcommand\sout{\bgroup \color{magenta} \ULdepth=-.5ex \ULset}

\begin{document}

\title{
Analysis of spectral modification of $\phi$ mesons at finite density using a transport approach in the \SI{12}{\giga\electronvolt} pA reactions
}


\newcommand\KEK {Present address: IPNS, KEK, Tsukuba 305-0801, Japan}
\newcommand\KEKacc {Present address: Accelerator Laboratory, KEK,  Tsukuba 305-0801, Japan}
\newcommand\KYOTOF {Present address: Institute for the Promotion of Excellence in Higher Education, Kyoto University,  Kyoto 606-8501, Japan}
\newcommand\NIAS {Present address: Institute for Innovative Science and Technology, Nagasaki Institute of Applied Science, Nagasaki, 851-0121, Japan}
\newcommand\ICEPP {Present address: International Center for the Elementary Particle Physics (ICEPP), University of Tokyo, Tokyo, 113-0033, Japan}
\newcommand\RCNP {Present address: Research Center for Nuclear Physics (RCNP), Osaka University, Ibaraki, 567-0047, Japan}
\newcommand\NAGOYA {Present address: Kobayashi-Maskawa Institute for the Origin of Particles and the Universe (KMI), Nagoya University, Nagoya 464-8602, Japan}
\newcommand\TOHOKU {Present address: Department of Physics, Tohoku University, Sendai 980-8578, Japan}
\newcommand\RIKENkaitaku {Present address: RIKEN Cluster for Pioneering Research, RIKEN, Wako 351-0198, Japan}
\newcommand\RIKENnishina {Present address: RIKEN Nishina Center, RIKEN, Wako 351-0198, Japan}
\newcommand\Deceased {Deceased}

\makeatletter
\renewcommand\maketitle{\par
   \begingroup

\renewcommand{\@fnsymbol}[1]{\ifcase##1\or \hbox{*}\or \titdagger\or a\or b\or c\or d\or e\or f\or g\or 
h\or i\or j\or k\or l  \else\@ctrerr\fi}
    
    \thispagestyle{titlepage}%
    \setcounter{footnote}{0}
    \renewcommand\thefootnote{\textsuperscript{\@fnsymbol\c@footnote}}%
    \def\@makefnmark{\rm\@thefnmark}%
    \long\def\@makefntext##1{
      {${\@thefnmark}$}{##1}}%
    \global\@topnum\z@   
    \@maketitle
      \markboth{\@shortauthorlist}{\@shorttitle}
    \@thanks
  \endgroup
  \@afterindentfalse
  \@afterheading
  \setcounter{footnote}{0}%
  \global\let\maketitle\relax
  \global\let\@maketitle\relax}
\makeatother

\author{
M.~Ichikawa$^{1,2}$\thanks{E-mail: michikaw@rcnp.osaka-u.ac.jp},
P.~Gubler$^1$,
J.~Chiba$^2$\thanks{\Deceased},
H.~En'yo$^3$,
Y.~Fukao$^{4,}$\thanks{\KEK},          
H.~Funahashi$^{4,}$\thanks{\KYOTOF},
H.~Hamagaki$^{5,}$\thanks{\NIAS},
M.~Ieiri$^2$,
M.~Ishino$^{4,}$\thanks{\ICEPP},
H.~Kanda$^{4,}$\thanks{\RCNP},
M.~Kitaguchi$^{4,}$\thanks{\NAGOYA},
S.~Mihara$^{4,\mathrm{a}}$, 
K.~Miwa$^{4,}$\thanks{\TOHOKU}$^{,\mathrm{a}}$,
T.~Miyashita$^4$, 
T.~Murakami$^4$, 
R.~Muto$^{4,}$\thanks{\KEKacc}, 
T.~Nakura$^4$,  
M.~Naruki$^4$,   
K.~Ozawa$^{5,\mathrm{a}}$,
F.~Sakuma$^{4,}$\thanks{\RIKENkaitaku},
O.~Sasaki$^2$,       
M.~Sekimoto$^2$,  
T.~Tabaru$^3$,   
K.~H.~Tanaka$^2$,    
M.~Togawa$^{4,\mathrm{a}}$,   
S.~Yamada$^{4,\mathrm{a}}$,      
S.~Yokkaichi$^3$,   
and Y.~Yoshimura$^4$
\authorcr
{\bf (KEK-PS E325 collaboration)}
}

\affil{
$^1${ Advanced Science Research Center, Japan Atomic Energy Agency,
Tokai 319-1195, Japan}\\
$^2${Institute of Particle and Nuclear Studies (IPNS), High Energy Accelerator Research Organization (KEK), Tsukuba, 305-0801, Japan}\\
$^3${RIKEN Nishina Center, RIKEN, Wako, 351-0198, Japan}\\
$^4${Department of Physics, Kyoto University, Kyoto 606-8502, Japan}\\
$^5${Center for Nuclear Study, Graduate School of Science, University of Tokyo, Tokyo 113-0033, Japan}\\
}



\vspace{-2mm}
\begin{abstract}%
The hadron spectrum at finite density is an important observable for exploring the origin of hadron masses.
In the KEK-PS E325 experiment, the di-electron decays of $\phi$ mesons inside and outside nuclei were measured using \SI{12}{\giga\electronvolt} pA reactions.
In the previous analysis, a significant excess was observed on the low-mass side of the $\phi$ meson peak in the data for slow-moving $\phi$ mesons ($\beta\gamma=p_{\phi}/m_{\phi}<1.25$) with the Cu target, and in-medium vector meson spectral modification was verified.
We newly employed the PHSD transport approach to take into account the time evolution of spatial density distribution of the target nuclei.
Consistent with the previous analysis, a significant excess was observed in the present analysis as well.
It was found that incorporating momentum dependence into the spectral modification leads to better agreement with the experimental results.
For the slow-moving $\phi$ mesons with the Cu target, the newly obtained modification parameters are consistent with those from the previous analysis within the uncertainties.
\end{abstract}
\subjectindex{D33}

\maketitle
\vspace{-5mm}
\section{Introduction}
\noindent
It is known that the masses of hadrons, particularly those of light hadrons, are much larger than the masses of their fermionic constituents, namely the current quarks.
Theoretically, quarks are believed to acquire larger constituent quark masses through their interactions with the non-trivial vacuum of the strong interaction, specifically, the quark condensate, which is an order parameter of the spontaneously broken chiral symmetry of Quantum Chromodynamics (QCD).

Conversely, the spontaneous breaking of chiral symmetry is predicted to be partially restored at high temperatures or finite densities.
Therefore, if experimental information on the quark condensate and hadron spectra in such environments can be obtained, a quantitative discussion on the relation between the quark condensate and hadron mass can be made.

The quantity of condensate was previously assessed by measuring the optical potential of pionic atoms some time ago \cite{Suzuki:2002ae}, and again recently \cite{nishi}.
Consequently, measuring the spectral modification of hadrons at finite densities is a crucial step toward elucidating the origin of hadron mass, clarifying whether the relation between hadron mass and chiral symmetry is indeed as direct as theoretically expected.

The KEK-PS E325 experiment measured the di-electron spectrum produced in \SI{12}{\giga\electronvolt} pC and pCu collisions within the light vector meson mass region \cite{muto}.
That analysis of the $\phi$ meson concluded that the data are consistent with a mass decrease of \SI{34}{\mega\electronvolt/\textit{c}^{2}} and a 3.6-fold increase in the decay width at normal nuclear density. 
However, Ref.~\cite{muto} did not incorporate the time evolution of the target nuclear density following the pA reaction into the analysis.
This issue can be addressed by using transport approaches, which allow one to trace the time evolution of the medium and its spatial density distribution and, in some cases, to treat off-shell effects of vector mesons \cite{Cassing:2009vt} and other particles \cite{Song:2020clw}.

For this reason, in the present work, we revisit the data obtained in the KEK-PS E325 experiment and reanalyze them using the PHSD transport approach.
The details of this approach relevant to the present study have already been discussed extensively in Refs.~\cite{Song:2022jcj,Gubler:2024ovg}.
The goal of this paper is to combine the results of Ref.~\cite{Gubler:2024ovg} with those of Ref.~\cite{muto} to determine which modification scenario for the $\phi$ meson at finite density best reproduces the experimental data.
In addition to providing a more detailed interpretation of the data, comparing the results of this work with the conclusions of Ref.~\cite{muto} offers valuable insights into the effects of assumptions made in the earlier analysis.

This paper is organized as follows.
Sect.~\ref{sec:phi} describes the advantages of using the $\phi$ meson to study its spectral modification at finite density.
Sect.~\ref{sec:e325} outlines the details of the KEK-PS E325 experiment.
Sect.~\ref{sec:phsd} presents the features of the PHSD transport approach, particularly those relevant to the present analysis.
The results of the analysis of the experimental data using PHSD are discussed in Sect.~\ref{sec:analysis} and \ref{sec:result}.
Finally, the paper is concluded in Sect.~\ref{sec:conclusion}.

\section{$\phi$ meson at finite density}
\label{sec:phi}
\noindent
The $\phi$ meson serves as a variable probe for investigating its spectrum at finite density. 
To begin with, the relation between light vector meson masses and quark condensates has been extensively studied using QCD sum rules \cite{sum_rule0,Asakawa:1994tp,Klingl:1997kf,Lee:1997zta,Zschocke:2002mn,Kampfer:2002pj,sum_rule1,Gubler:2015yna,Gubler:2016itj,Kim:2019ybi,Kim:2022eku}.
In addition, hadronic effective models have been employed to compute the $\phi$ meson spectral function at finite density \cite{Klingl:1997tm,Cabrera:2002hc,Gubler:2015yna,Gubler:2016itj,Cabrera:2016rnc}, providing---although model dependent---information about its mass and decay width as functions of density.

Moreover, when measuring the mass spectra of hadrons in nuclei experimentally, final-state interaction between the decay products and nuclei can pose a problem.
However, this effect is suppressed for vector mesons decaying into di-leptons, as they do not interact strongly with the medium.
Although the $\rho$ and $\omega$ mesons are also light vector mesons, they have proven to be less ideal probes for studying spectral modifications at finite density.
The $\rho$ meson has a large decay width of \SI{147.4}{\mega\electronvolt/\textit{c}^{2}}, making it difficult to accurately separate possible mass change and width broadening.
Meanwhile, the $\omega$ meson mass is very close to that of the $\rho$ meson, which makes it difficult to solve an entanglement on the mass spectrum.
In contrast, the $\phi$ meson has a narrow width and no nearby resonance, facilitating a more definitive and quantitative discussion.

The interaction between $\phi$ mesons and nucleons or nuclei has been investigated in various experimental and theoretical studies. 
In particular, the correlation function of p$\phi$ has recently been measured using femtoscopy by the ALICE experiment \cite{corr}.
Combined with a lattice QCD calculation by the HAL QCD Collaboration, these results provide an indication of a bound spin-1/2 p$\phi$ system \cite{halqcd0,halqcd1}.
However, mainly due to the small cross section, few experiments have addressed the spectral shape and its modification of the $\phi$ meson at finite density.
Although the CLAS-g7 experiment at Jefferson Lab measured the $e^{+}e^{-}$ spectrum from $\gamma$A reactions, the discussion of spectral modification was limited to the $\rho$ meson only \cite{clas}.
The NA60 experiment at CERN-SPS measured the $\mu^{+}\mu^{-}$ spectrum from In-In collisions at \SI{158}{A\giga\electronvolt}.
Although a width broadening of the $\rho$ meson was observed \cite{na60_0}, no significant spectral modification was observed for the $\phi$ meson \cite{na60_1}.

To date, the KEK-PS E325 experiment \cite{muto} is the only experiment which observed the spectral modification of the $\phi$ meson in the invariant mass spectrum.

\section{\phiee\ spectra measured in the KEK-PS E325 experiment}
\label{sec:e325}
\noindent
A significant spectral modification in the \phiee\ invariant mass spectra was observed using \SI{12}{\giga\electronvolt} proton beam and nuclear targets of C and Cu in the 2001-2002 data-taking run \cite{muto}.

In the analysis, the spectra obtained in the experiment were classified by target nucleus, namely C and Cu, and three $\beta\gamma$ ($=p_{\phi}/m_{\phi}$) regions of the $\phi$ mesons in the laboratory frame.
Since the typical flight length of $\phi$ mesons is large relative to the size of the nucleus, most of the produced $\phi$ mesons decay outside the target nuclei.
Therefore, if there is any modification of the $\phi$ meson spectral function in dense matter, the magnitude of the modification is expected to be larger for slower $\phi$ mesons and for larger target nucleus.

The observed data were compared with the resonance shapes and a significant difference was observed in the slowest $\beta\gamma$ region ($\beta\gamma<1.25$) for the Cu target.
It is consistent with the picture described above.
In the analysis, the original resonance shape was the non-relativistic Breit--Wigner function:
\begin{equation}\label{eq:nrbw}
    A(m)=\frac{1}{2\pi}\frac{\Gamma_{\phi}}{(m-m_{\phi})^{2}+(\frac{\Gamma_{\phi}}{2})^{2}},
\end{equation}
and the isotropic decay into \ee\ pair in the rest frame of the $\phi$ meson was assumed.
Then, the internal radiative correction (IRC) was applied and the $e^{+}$ and $e^{-}$ were fed into the detailed detector simulation
using the Geant4 toolkit \cite{geant0, geant1, geant2} to take account of the experimental effects, for example, the energy loss in the target and detector material, the detector acceptance and tracking performance.

Moreover, to compare the data with theoretical predictions
on spectral modification in dense matter,
a parametrization of the Breit--Wigner shape as Eq.~(\ref{eq:nrbw}) was performed using two parameters $k_{1}$ and $k_{2}$ as:
\begin{equation}\label{eq:shift}
    m_{\phi}(\rho)=(1-k_{1}\frac{\rho}{\rho_{0}})m_{0},
\end{equation}
\begin{equation}\label{eq:broad}
    \Gamma_{\phi}(\rho)=(1+k_{2}\frac{\rho}{\rho_{\text{0}}})\Gamma_{0},   
\end{equation}
where $\rho$ is the local baryon density at the position of $\phi$ meson, $m_{0}$ and $\Gamma_{0}$ are its vacuum mass and decay width, $\rho_{0}$ is the normal nuclear density.

To make a spectrum depending on the density at the decay point of $\phi$ mesons, a Monte Carlo-type model calculation was performed as follows.
The production point of $\phi$ was assumed to be distributed in the target nucleus, proportionally to the nucleon density of the nucleus. Woods--Saxon type density distribution was assumed:
\begin{equation}\label{eq:ws}
    \rho(r)=\frac{N}{1+\exp(\frac{r-R}{\tau})}\rho_{0},
\end{equation}
where $r$ is the distance from the center of the nucleus, $R$ is the half density radius of the nucleus, $\tau$ is related to the surface thickness, and $N$ is a normalization factor.
The momentum distribution of $\phi$ was generated using the code JAM \cite{jam}, which approximately reproduces the observed momentum distribution.
The possible correlation of the production point and the initial momentum was neglected.
The flight of the $\phi$ mesons was simulated step-by-step with a length of \SI{0.1}{\femto\meter}, conserving their initial three momentum when the mass and width are changed in each step depending on the density.

In each	step, the decay probability was calculated using $\beta\gamma c \tau$ depending on the mass and width at the middle point of the step.
If meson was decayed in the step, the mass is calculated following the probability distribution of the Breit--Wigner form with the mass and width in the step.

Once the mass distribution was calculated for a pair of $k_{1}$ and $k_{2}$, the same procedures as above mentioned for the Breit--Wigner
distribution, namely, the isotropic decay into \ee\ pair, IRC, and the detector simulation were applied and a spectrum was obtained to fit the experimental data.
On the density-dependence of the partial width decaying to \ee\, two extreme cases were examined: a density-dependent case, in which the partial decay width scales proportionally to the total width, and a density-independent case.
The partial width, unlike the total width, does not directly lead to a broadening of the spectral shape.
Nevertheless, its increase at finite density could enhance the $e^{+}e^{-}$ decay probability, which in turn influences the spectral shape to be observed.

Many pairs of $k_{1}$ and $k_{2}$ were examined	to fit the data, and the best-fit parameters were obtained as $k_{1}=0.034$ and $k_{2}=2.6$ for the density-dependent partial width.

However, a potentially important aspect of the pA reaction---which was not considered in the analysis of Ref.~\cite{muto}---is that the target nucleus no longer retains its nuclear form, leading to a change in the spatial density distribution. 
If the time scale of this deformation is comparable to that of the production and decay of $\phi$ mesons, its effect should be taken into account.
The production point of $\phi$ mesons may also deviate from a form proportional to the spatial distribution of the target density in vacuum.
This is a simple assumption reflects the mass-number dependence of the $\phi$ meson production cross section as $\sigma(A)\propto A^{1}$, which was measured by the experiment \cite{tabaru}.
To take these effects into account, we employ in this work the PHSD transport approach, which can simulate the dynamics of all participating particles in the pA reaction.

\section{PHSD transport approach}
\label{sec:phsd}
\noindent
The Parton-Hadron-String Dynamics (PHSD) is an off-shell transport approach developed to describe interactions among hadrons and partons.
It is formulated using the generalized off-shell Cassing--Juchem transport equations.
PHSD has been widely applied to relativistic heavy-ion collisions and has successfully reproduced.
For example, the rapidity distributions, transverse mass spectra, and the centrality of the reaction and transverse momentum dependence of the eliptic flow of charged hadrons in Au + Au collisions at $\sqrt{s}=200$~\si{\giga\electronvolt} at RHIC \cite{phsd_rhic}, as well as the di-lepton yields in In + In collisions at \SI{158}{A~\giga\electronvolt} at the NA60 experiment \cite{phsd_sps}.

Simulations using PHSD to study the production of $\phi$ mesons in heavy-ion collisions and pA reactions have been conducted in Refs.~\cite{Song:2022jcj,Gubler:2024ovg}.
We refer interested readers to these references for more details on the PHSD approach, and focus here only on the features and issues relevant to the present study.

\subsection{Time integration method}
\noindent
Because of the relatively small branching ratio of the di-lepton decay of vector mesons, generating the di-lepton spectrum with sufficient statistics is a challenging task for transport approaches.
To overcome this problem, the time integration (or shining) method has proven to be a useful tool. 
In this method, a virtual $\phi\rightarrow e^{+}e^{-}$ decay is assumed to occur at each time step of the simulation, allowing statistics to be accumulated equal to the number of generated $\phi$ mesons times the number of time steps. 
This prescription is implemented using the di-electron decay probability at each time step, which is given in Sect.~\ref{sec:generator}.
Note that the $\phi$ meson in this method never actually decays into a di-lepton pair, but only into hadronic decay channels.
This approximation can be justified by the smallness of the di-lepton branching ratio.

\subsection{Density dependence of lifetime and width}
\label{sec:lifetime}
\noindent
While the $\phi$ meson lifetime in PHSD decreases with increasing density, the reduction is indirectly implemented through absorptions with other hadrons during the simulation process.
Consequently, the lifetime emerges from the complex interplay of the absorption cross sections of these hadronic processes.
On the other hand, the present analysis requires $\phi$ meson spectra to be obtained for various values of $k_{2}$ that is, for different strength of the density dependence of the lifetime.
Therefore, it is necessary to account for the lifetime reduction originally included in PHSD through hadronic absorption processes.
The details of our strategy for addressing this issue are discussed in Sect.~\ref{sec:generator}.

\subsection{Stability of target nucleus}
\label{sec:stability}
\noindent
It is generally known that transport approaches have difficulty conserving the initial density distribution of nuclei as time progresses (see, for instance, the discussion in Ref.~\cite{TMEP:2016tup}).
In PHSD, this problem is partially addressed by freezing the motion of nucleons inside the target nucleus until the projectile hits, and by introducing a mean-field potential, which also helps the target nucleus retain its initial shape.
Nevertheless, PHSD still shows that the target nuclei slightly increase in size over time, even in cases where the projectile misses the target, which should therefore not alter its structure. 
The impact of this artificial deformation is estimated in Sect.~\ref{sec:error}.

\section{Analysis}
\label{sec:analysis}
\noindent
In the analysis, the data points in the six mass spectra to be analyzed, the parameterization of the spectral modification, and the treatment of density dependence of the partial width of the di-electron decay were all performed in the same manner as in Ref.~\cite{muto}.
The di-electron spectra for each target nucleus (C and Cu data, respectively) were divided into three $\beta\gamma$ regions: $\beta\gamma<1.25$ (slow), $1.25\leq\beta\gamma<1.75$ (middle) and $\beta\gamma\geq1.75$ (fast).
The spectral modification was parameterized using Eqs.~(\ref{eq:shift})~and~(\ref{eq:broad}).
For the partial width of the di-electron decay, two extreme scenarios were considered: one in which it depends on the density through the same parameter $k_{2}$ as the total width, and another in which it remains constant, independent of the density.

Let us summarize here the steps needed to simulate the di-electron invariant mass spectrum for various $\phi$ meson modification scenarios.
First, the momentum of the $\phi$ mesons at the time their decay, as well as the baryon density at their decay point and time, are extracted from our PHSD transport approach of the \SI{12}{\giga\electronvolt} pA reaction. 
Next, the momentum and density data are used to generate $\phi$ mesons decaying into di-electrons, and is then further modified by incorporating the IRC effects. Next, electron and positron are fed to the detector simulation to take account of the experimental effect including energy loss in the target and detector materials and also the position resolution.
Then, the invariant mass is calculated for each electron-positron pair and used to fit the experimental data with the background shape, an exponential function, $A\exp(-B(m-C))$, where $m$ is the mass.
The parameters of the exponential function are obtained by fitting each experimental spectrum while excluding the signal region (\SI{0.95}{\giga\electronvolt/\textit{c}^{2}}--\SI{1.05}{\giga\electronvolt/\textit{c}^{2}}).
A grid search is performed over the $k_{1}$-$k_{2}$ parameter space to obtain the set of values that provides the best agreement with the experimental data of Ref.~\cite{muto}.

\subsection{$\phi$ meson generator}
\label{sec:generator}
\noindent
Here, we outline how we construct the $\phi$ meson generator based on the PHSD transport approach of the pA reaction.
The transport approach provides a wide range of kinematical and other information (such as energy, momentum, position, and local density at that position) at each time step.
From these data, joint distributions of both the local density and momentum components parallel and perpendicular to the beam ($p_{\text{z}}$ and $p_{\text{T}}$, respectively) can be obtained. 
In this process, two types of weight must be considered at each step.

The first is the di-electron decay probability of the $\phi$ mesons at each step, $w^{\phi\rightarrow e^{+}e^{-}}$, which is expressed as
\begin{equation}\label{eq:ee_decay_prob}
w^{\phi\rightarrow e^{+}e^{-}}(\rho,\gamma,\Delta t)
=1-\exp(-\frac{\Gamma^{ee}(\rho)\Delta t}{\hbar c\gamma}),
\end{equation}
where $\rho$ is the local density at the $\phi$ meson's position, $\gamma$ is the Lorentz factor, $\Delta t$ is the time interval between simulation steps, and $\Gamma^{ee}$ is the di-electron decay width.
In the present analysis, both the case where $\Gamma^{ee}$ is a constant and the case where it depends on density as
\begin{equation}
    \label{eq:ee_width_rho_dep}
    \Gamma^{ee}(\rho)=(1+k_{2}\frac{\rho}{\rho_{0}})\Gamma^{ee}_{0},
\end{equation}
are considered.
Here, the value of $\rho_{0}$ used is \SI{0.168}{\femto\meter^{-3}}, and $\Gamma^{ee}_{0}$ is the partial decay width in vacuum, equal to \SI{1.266e-6}{\giga\electronvolt/\textit{c}^{2}}.
This value is calculated as the product of the total width, \SI{4.249e-3}{\giga\electronvolt/\textit{c}^{2}}, and the di-electron branching ratio, \num{2.979e-4} \cite{pdg}.

\begin{figure}[!h]
\centering
\includegraphics[width=170mm]{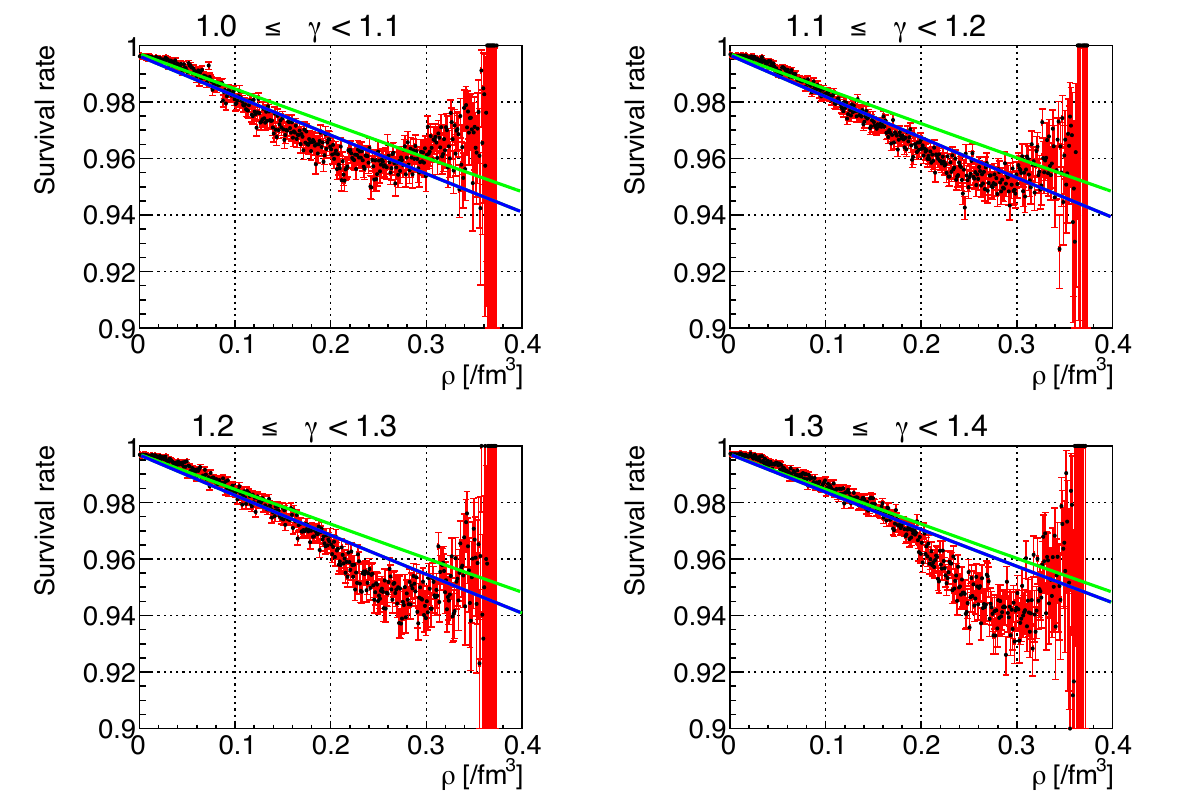}
\caption{
    Examples of the density dependence of the survival rate for each Lorentz factor.
    Black points with error bars represent the survival rates of each Lorentz factor and density.
    Red lines show Eq.~(\ref{eq:survival_rate}) with $k_{2}^{\text{PHSD}}=7.18$.
    Green lines show the fitting results using Eq.~(\ref{eq:survival_rate}) with $k_{2}^{\text{PHSD}}$ as a free parameter.
}
\label{fig:phsd_k2_samples}
\end{figure}

\begin{figure}[!h]
\centering
\includegraphics[width=100mm]{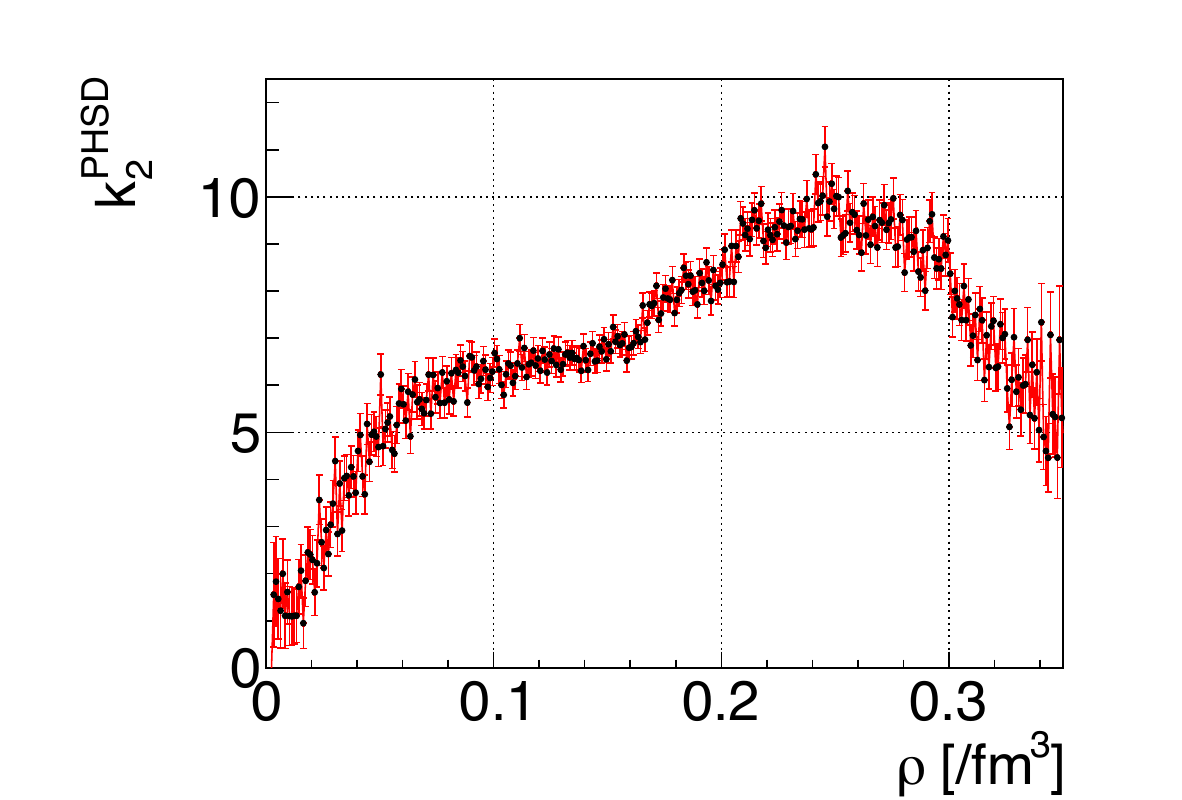}
\caption{
    Density dependence of $k_{2}^{\text{PHSD}}$ averaged over all Lorentz factors.
    Black points with red error bars indicate $k_{2}^{\text{PHSD}}$ values for each density bin.
    The density region above \SI{0.35}{\femto\meter^{-3}} is excluded due to insufficient statistics.
}
\label{fig:rho_phsd_k2}
\end{figure}

The second factor to consider is the correction of the survival rate due to absorption reactions included in the PHSD transport approach, as explained in Sect.~\ref{sec:lifetime}.
First, we must determine how much the $\phi$ meson decay width increases due to these absorption processes.
For this purpose, the $\phi$ meson survival rate at each time step is expressed as
\begin{equation}
\label{eq:survival_rate}
p(\Gamma, \Delta t, \gamma)=\exp{(-\frac{\Gamma\Delta t}{\hbar c\gamma})}.
\end{equation}
By applying Eq.~(\ref{eq:broad}) to $\Gamma$ in Eq.~(\ref{eq:survival_rate}) and fitting the survival rates extracted from the PHSD simulation, as shown in Fig.~\ref{fig:phsd_k2_samples} using $k_{2}$ as a free parameter, the intrinsic value of $k_{2}$ in PHSD, denoted as $k_{2}^{\text{PHSD}}$, was obtained.
As shown in Fig.~\ref{fig:rho_phsd_k2}, this value was found to be 7.18.
When varying the lifetime of the $\phi$ meson according to the value of $k_{2}$, it is necessary to take into account the influence of $k_{2}^{\text{PHSD}}$.
We correct for this mismatch using the following weighting factor:
\begin{equation}\label{eq:survival_rate_correction}
\begin{split}
w^{corr.}(\rho,\gamma,\Delta t)
=\prod\frac{p^{\text{decay}}(\rho,\gamma,\Delta t)}{p^{\text{decay,PHSD}}(\rho,\gamma,\Delta t,)} \\
=\prod\exp(\frac{(\Gamma^{\text{PHSD}}(\rho)-\Gamma(\rho))\Delta t}{\hbar c\gamma}) \\
=\prod\exp(\frac{(k_{2}^{\text{PHSD}}-k_{2})\frac{\rho}{\rho_{0}}\Gamma_{0}\Delta t}{\hbar c\gamma}), 
\end{split}
\end{equation}
where $p^{\text{decay}}$ is the probability of $\phi$ meson decay assuming a decay width determined by $k_{2}$ in Eq.~(\ref{eq:broad}), $p^{\text{decay, PHSD}}$ is the respective decay probability of total decay of $\phi$ mesons calculated from the PHSD output based on the absorption processes mentioned above, while $\Gamma$ and $\Gamma^{\text{PHSD}}$ represent the corresponding decay widths.

When generating the four momenta of the $\phi$ mesons, the local densities are first randomly selected in proportion to the relevant density distribution obtained from our PHSD simulation.
The mass is sampled from the non-relativistic Breit--Wigner distribution described in Eq.~(\ref{eq:nrbw}), with the resonance mass and width calculated from Eqs.~(\ref{eq:shift})~and~(\ref{eq:broad}) at the obtained density.
Although PHSD is capable of simulating the time evolution of $\phi$ mesons while accounting for their in-medium spectral modification, in the present analysis, the spectral modification was applied after the PHSD calculation in order to reduce computing cost.
$p_{\text{T}}$ and $p_{\text{z}}$ are determined according to the momentum correlations corresponding to the selected density, and the momenta are randomly rotated around the z-axis.
Finally, the mass is sampled according to Eq.~(\ref{eq:nrbw}).

\subsection{Internal radiative corrections and experimental effects}
\begin{figure}[!h]
\centering
\includegraphics[width=170mm]{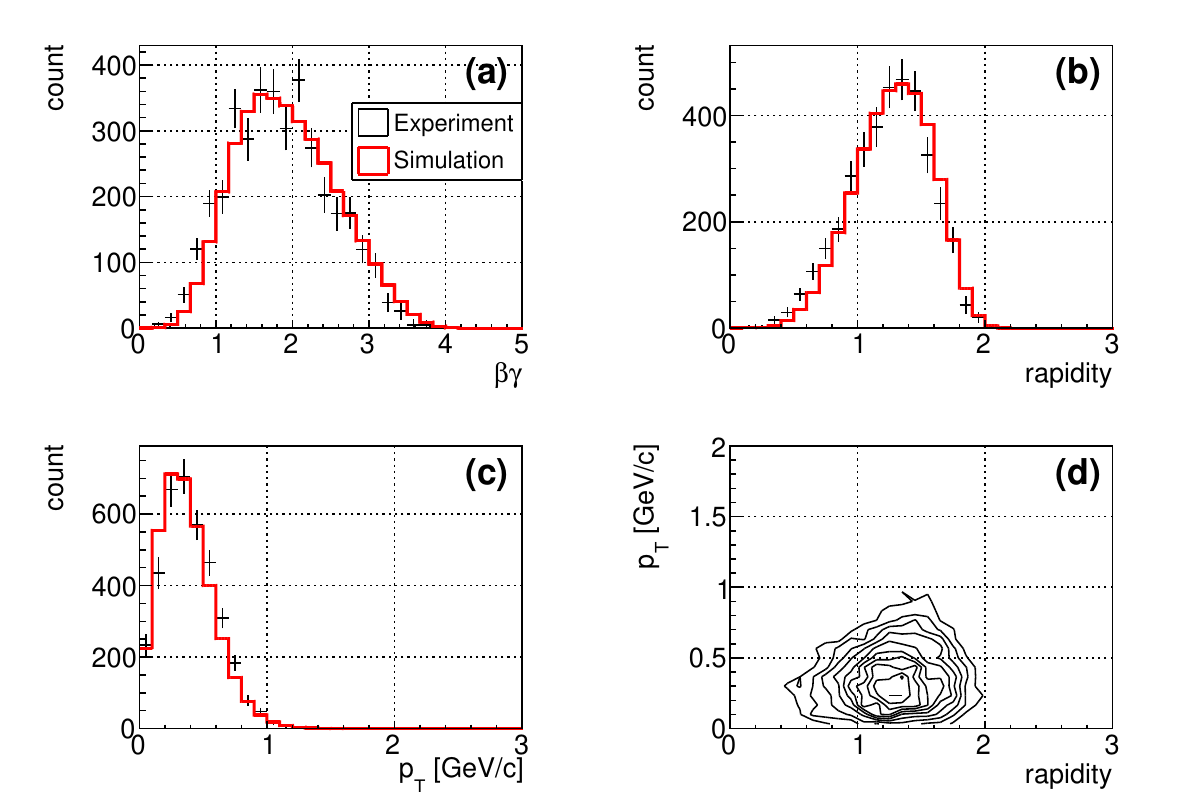}
\caption{
    Kinematical distribution of $\phi$ mesons, as functions of (a) $\beta\gamma$, (b) rapidity, (c) transverse momentum, and (d) transverse momentum versus rapidity.
    The black lines with error bars represent real data and red lines represent the simulation results as described in Sect.~\ref{sec:analysis}.
}
\label{fig:mom}
\end{figure}

\noindent
Internal radiative corrections, namely, vertex correction, vacuum polarization, and internal Bremsstrahlung due to electromagnetic interactions, are evaluated using PHOTOS 3.64; a Monte Carlo simulation code for QED radiative corrections \cite{photos}. 
Examples of the effect of IRC on the invariant mass spectrum of $\phi$ meson are provided in Ref.~\cite{ichikawa_phsd}.
Experimental effects are evaluated using the detector simulation as same as described in Sect.~\ref{sec:e325}.
Figure~\ref{fig:mom} shows the kinematical distributions of the $\phi$ mesons obtained in the analysis and those measured in the KEK-PS E325 experiment.
As can be seen in this figure, the experimental momentum distributions are well reproduced by the present study.

\section{Result and discussion}
\label{sec:result}
\subsection{Fit result using PHSD}
\begin{figure}[!h]
\centering
\includegraphics[width=170mm]{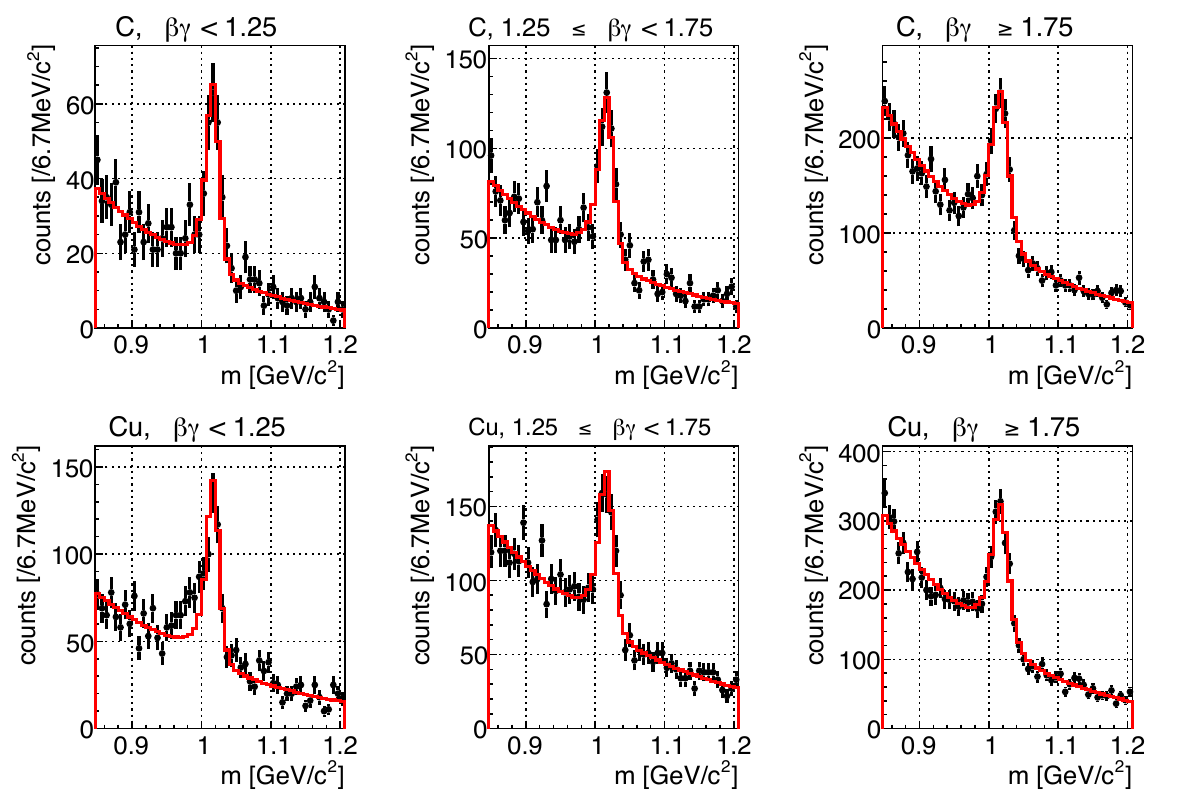}
\caption{
    Experimental data and fit results assuming no spectral modification.
    The black points with error bars represent the experimental data of Ref.~\cite{muto}.
    The red lines show the fit results.
}
\label{fig:fit_no_modify}
\end{figure}

\noindent
First, the results of fitting the experimental data with spectra generated according to the analysis described in Sect.~\ref{sec:analysis}, for the no mass-modification case ($k_{1}=0, k_{2}=0$), are shown in Fig.~\ref{fig:fit_no_modify}.
The minimum $\chi^{2}$ values for each spectrum are listed in Table~\ref{tab:no_modification}.
\begin{table}[t]
\centering
\caption{
    Minimum $\chi^{2}$ values for each spectrum without spectral modification.
}
\label{tab:no_modification}
\begin{tabular}{ccc}
\hline
$\beta\gamma$ & $\chi^{2}_{\text{min}}/\text{ndf}$ (C) & $\chi^{2}_{\text{min}}/\text{ndf}$ (Cu) \\
\hline
Slow   & 38/53 & 93/53 \\
Middle & 68/53 & 47/53 \\
Fast   & 46/53 & 54/53 \\
\hline
\end{tabular}
\end{table}
Notably, for the Cu data in the slow $\beta\gamma$ region, the $\chi^{2}/\text{ndf}$ value is as large as 93/53, which allows the no-modification hypothesis to be rejected at the 99\% confidence level.

\begin{figure}[!h]
\centering
\includegraphics[width=170mm]{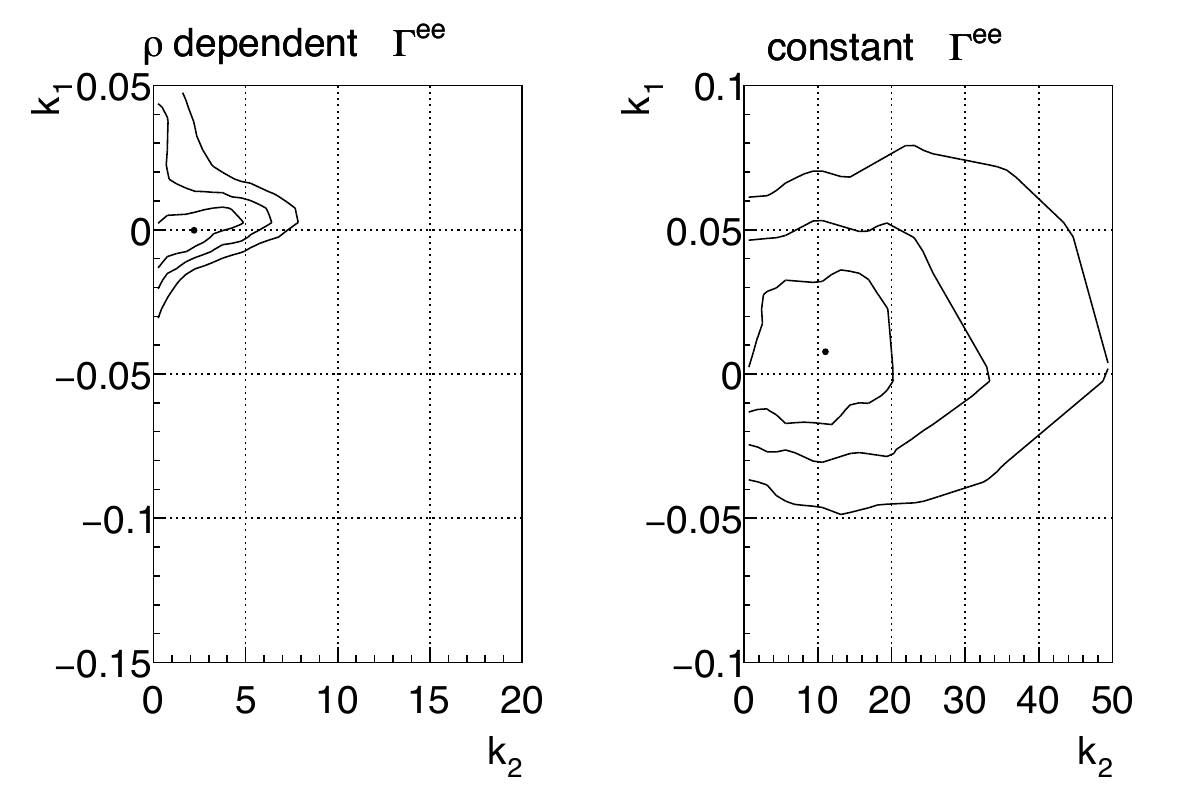}
\caption{
    $\chi^{2}$ distribution in the $k_{1}$-$k_{2}$ parameter space.
    The contours represent $1\sigma$ ($\Delta\chi^{2}=2.30$), $2\sigma$ ($\Delta\chi^{2}=6.18$) and $3\sigma$ ($\Delta\chi^{2}=11.83$) confidence levels for the two parameters. They were obtained by interpolating the values at the grid points.
    Black points indicate the minimum $\chi^{2}$ points determined by the fit.
}
\label{fig:k1k2_scan}
\end{figure}

\begin{figure}[!h]
\centering
\includegraphics[width=170mm]{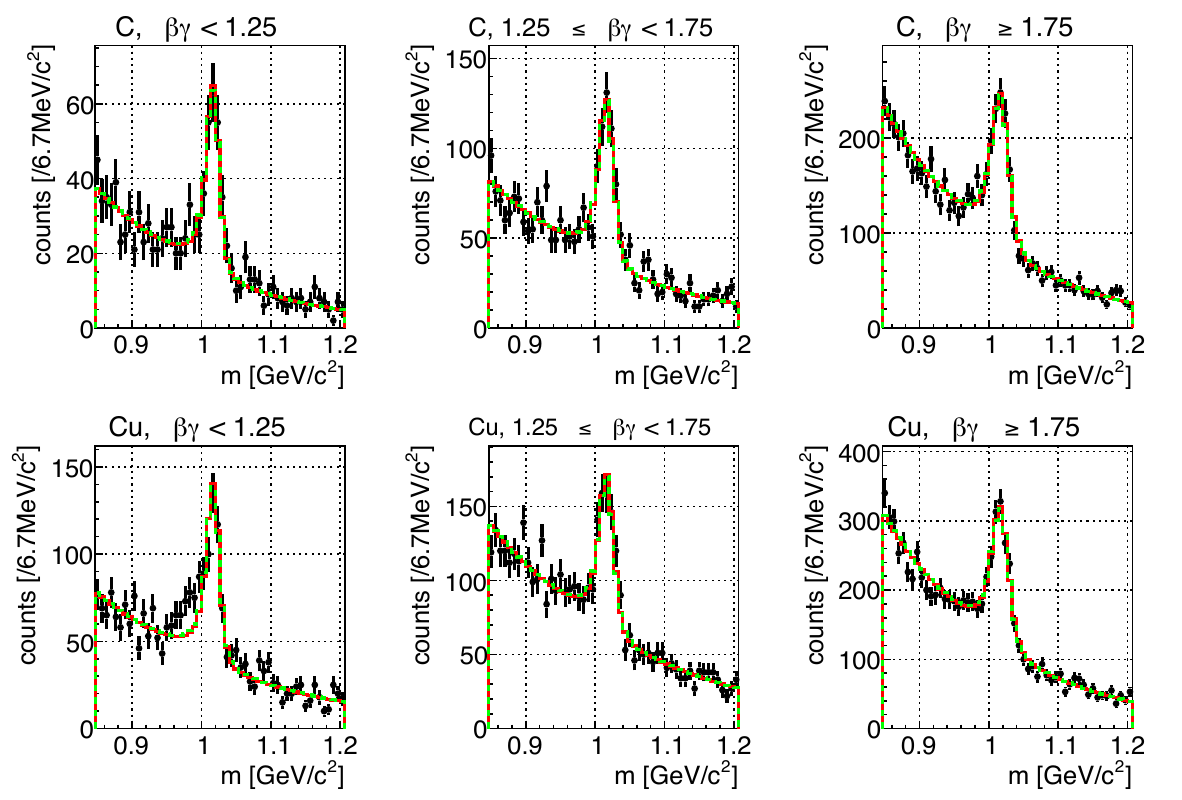}
\caption{
    Experimental data and fit results assuming spectral modification.
    For the fitting, spectra corresponding to the simulated grid point ($k_{1}$, $k_{2}$) nearest to the values in Table~\ref{tab:k1k2_common} were used.
    Black points with error bars show the data.
    Red lines show the fit results assuming density-dependent partial widths, while green dashed lines represent the fit results for constant partial widths.
}
\label{fig:common_param_fit_result}
\end{figure}

Next, the results of the $k_{1}$ and $k_{2}$ scans for the density-dependent and constant partial width cases are shown in Fig.~\ref{fig:k1k2_scan}.
The analysis described in Sect.~\ref{sec:analysis} was performed for various grid points in the parameter space ($k_{1}$, $k_{2}$), and the corresponding $\chi^{2}$ values were calculated at each point.
To obtain the best-fit values, quadratic functions were fitted to the grid points near the minimum $\chi^{2}$ value.
The resulting best-fit parameters are shown in Fig.~\ref{fig:common_param_fit_result} and summarized in Table~\ref{tab:k1k2_common}, where common values of $k_{1}$ and $k_{2}$ were applied across all target nuclei and $\beta\gamma$ regions.
\begin{table}[t]
\centering
\caption{
    Fit results assuming common $k_{1}$ and $k_{2}$ for all nuclei and $\beta\gamma$ regions.
    Errors are not included.
}
\label{tab:k1k2_common}
\begin{tabular}{ccccccc}
\hline
Partial decay width & $k_{1}$ & $k_{2}$ & $\beta\gamma$ &
$\chi^{2}$ (C) & $\chi^{2}$ (Cu) & $\chi^{2}/\text{ndf}$ (total) \\
\hline
Density dependent & 0.000 & 2.2 &
    Slow   & 38 & 89 & \\
&&& Middle & 68 & 46 & \\
&&& Fast   & 46 & 56 & \\
&&& Total  &    &    & 343/316 \\
\hline
Constant & 0.008 & 11.0 &
    Slow   & 38 & 86 & \\
&&& Middle & 69 & 46 & \\
&&& Fast   & 47 & 57 & \\
&&& Total  &    &    & 342/316 \\
\hline
\end{tabular}
\end{table}
The p-values for the density-dependent and constant partial width cases are 14\% and 15\%, respectively.
Thus, both cases are not rejected statistically.
However, the fit for the Cu data in the slow $\beta\gamma$ region still fails to reproduce the data satisfactorily.
The corresponding $\chi^{2}$ values are 89 and 86, respectively (with 54 bins), indicating that assuming a common set of $k_{1}$ and $k_{2}$ values for all spectra may not capture the underlying physics.

\begin{figure}[!h]
\centering
\includegraphics[width=170mm]{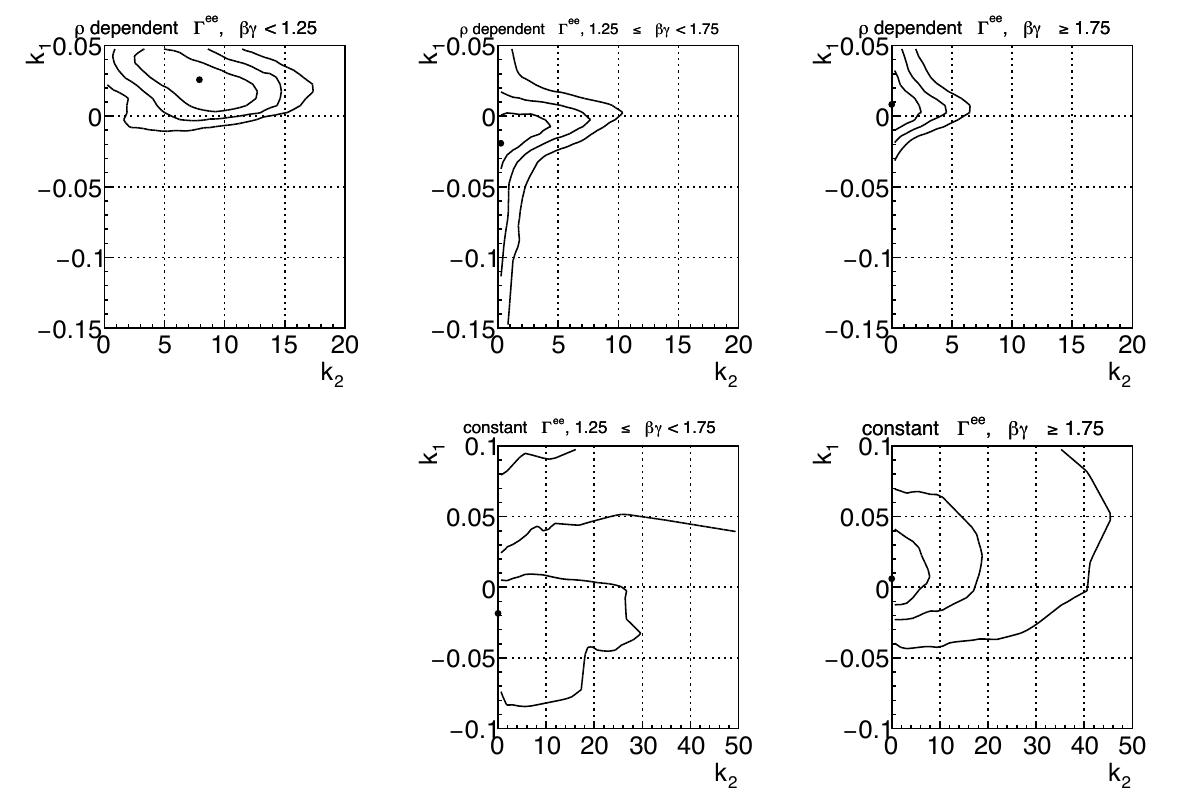}
\caption{
    $\chi^{2}$ distribution in the $k_{1}$-$k_{2}$ parameter space for each $\beta\gamma$ region.
    The contour definitions are the same as in Fig.~\ref{fig:k1k2_scan}.
    Black points indicate the minimum $\chi^{2}$ values obtained from quadratic fits.
    The panel corresponding to the slow $\beta\gamma$ region in the constant partial width case is excluded due to unphysically large best-fit values of $(k_{1},k_{2})$.
}
\label{fig:k1k2_scan_bg_dep}
\end{figure}

\begin{figure}[!h]
\centering
\includegraphics[width=170mm]{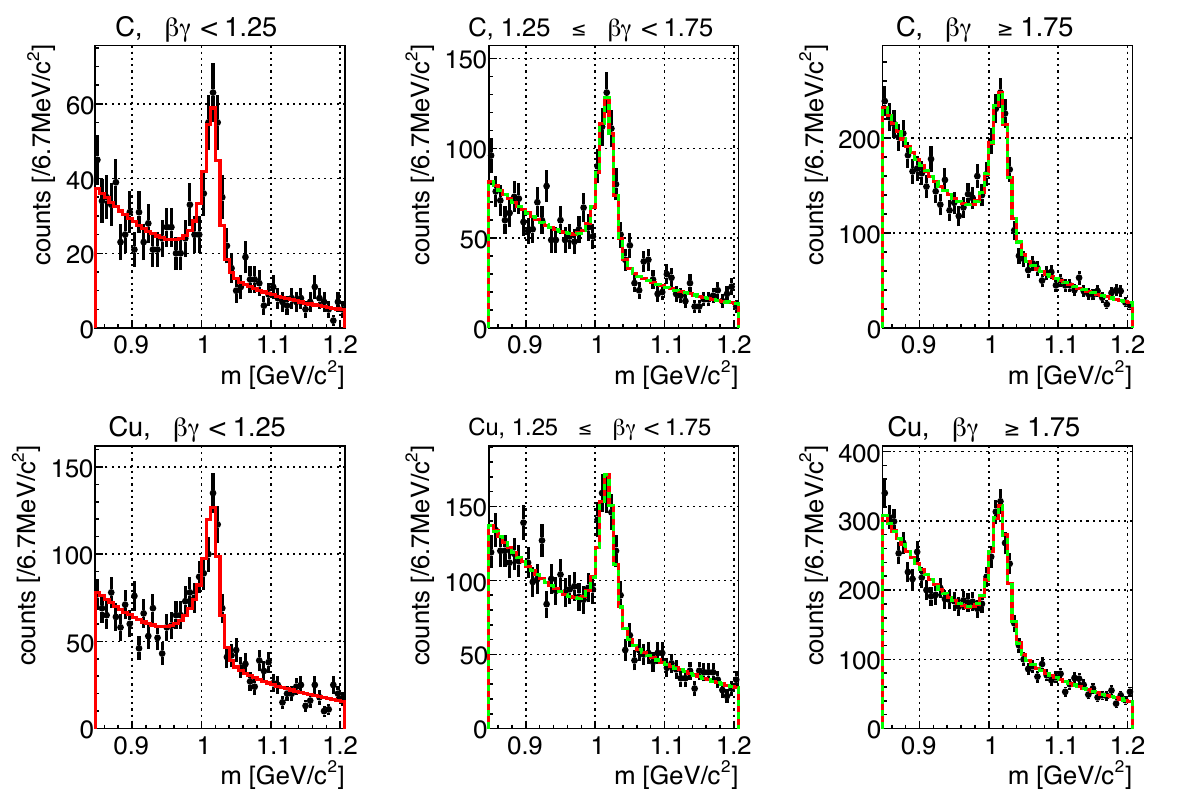}
\caption{
    Experimental data and fit results assuming spectral modifications with $\beta\gamma$ dependence.
    For the fit, the spectra at the simulated grid points closest to the values in Table~\ref{tab:k1k2_betagamma} were used.
    Black points with error bars represent the experimental data of Ref.~\cite{muto}. 
    Red lines show the fit results for the density-dependent partial width case.
    Green dashed lines show the fit results for the constant partial width case.
    No spectra are shown for the constant partial width case in the slow regions of C and Cu due to unphysical $k_{1}$ and $k_{2}$ values.
}
\label{fig:each_param_fit_result}
\end{figure}

To explore possible momentum dependence, we allow $k_{1}$ and $k_{2}$ to vary by $\beta\gamma$ region, keeping them common across targets.
Momentum-dependent modifications of the $\phi$ meson pole mass and decay width have been proposed in, for example, Refs.~\cite{Lee:1997zta,Kim:2022eku,Cabrera:2016rnc}.
The results are represented in Fig.~\ref{fig:k1k2_scan_bg_dep} and Table~\ref{tab:k1k2_betagamma}.
\begin{table}[t]
\centering
\caption{
    Fit results assuming common values of $k_{1}$ and $k_{2}$ for both C and Cu targets, evaluated separately for each $\beta\gamma$ region.
    Errors are not shown.
    Unphysical region with $k_{2}<0$ are excluded.
}
\label{tab:k1k2_betagamma}
\begin{tabular}{ccccccc}
\hline
Partial decay width & $\beta\gamma$ & $k_{1}$ & $k_{2}$ &
$\chi^{2}$ (C) & $\chi^{2}$ (Cu) & $\chi^{2}/\text{ndf}$ (total) \\
\hline
Density dependent &
  Slow   & 0.026   & 7.9 & 42 & 73 & 115/104 \\
& Middle & --0.019 & 0.2 & 66 & 47 & 112/104 \\
& Fast   & 0.008   & 0.0 & 45 & 55 & 100/104 \\
\hline
Constant &
  Slow   & - & - & - & - & - \\
& Middle & --0.018 & 0.0 & 66 & 46 & 112/104 \\
& Fast   & 0.006   & 0.0 & 45 & 55 & 100/104 \\
\hline
\end{tabular}
\end{table}
In this analysis, the common $k_{1}$ and $k_{2}$ values are used for both C and Cu data within each $\beta\gamma$ region, allowing for an approximate evaluation of the momentum dependence of the $\phi$ meson spectral properties.
In the constant partial width case, the best-fit values of $k_{1}$ and $k_{2}$ for the slow $\beta\gamma$ region become unphysically large ($k_{1}>0.5$, $k_{2}>100$). 
The corresponding best-fit spectra are shown in Fig.~\ref{fig:each_param_fit_result}.
Compared to Fig.~\ref{fig:common_param_fit_result}, the fit to the Cu data in the slow $\beta\gamma$ region is clearly improved in Fig.~\ref{fig:each_param_fit_result}.

\subsection{Systematic uncertainty}
\label{sec:error}
\noindent
In this section, we discuss the estimation of systematic uncertainties associated with the PHSD simulation, in particular, we consider the effects of the instability of the target nucleus and the intrinsic lifetime shortening characterized by the parameter $k_{2}^{\text{PHSD}}$.
In addition, we evaluate the impact of the choice of functional form and parameter range of the background shape in the spectral fit.

\subsubsection{Stability of target nucleus}
\begin{figure}[!h]
\centering
\includegraphics[width=170mm]{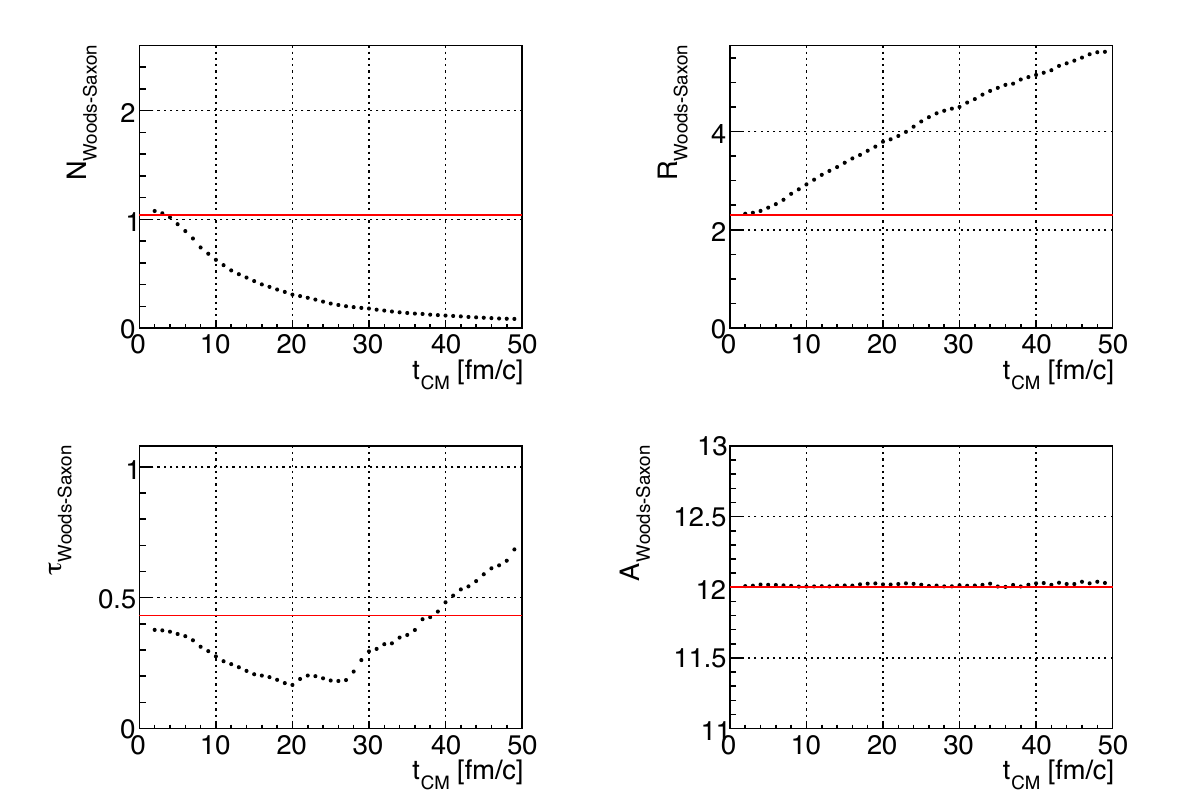}
\caption{
    Time evolution of the Woods--Saxon parameters (as defined in Eq.~(\ref{eq:ws})).
    $t_{CM}$ represents the time in the center-of-mass frame.
    The top two panels and bottom-left panel show the parameters directly, while the bottom-right panel shows the corresponding effective mass number of the $^{12}$C target.
    The red lines indicate the static Woods--Saxon parameters for a $^{12}$C nucleus.
}
\label{fig:t_dep_of_ws}
\end{figure}

\noindent
As described in Sect.~{\ref{sec:stability}}, the spatial nucleon distribution of the target nucleus in PHSD simulations deviates from its initial Woods--Saxon form due to two main effects:
\begin{enumerate}
    \item{Physical effects arising from interactions between the projectile and the target, such as a local density enhancement near the collision point or spatial expanding of the nucleon distribution due to energy deposition.}
    \item{Artificial expanding of the nucleon distribution in the target nucleus caused by unintended instability in the transport approach.}
\end{enumerate}
Effect~(1) is physical and should be properly incorporated into the analysis, while Effect~(2) is unphysical and ideally should be minimized or corrected.
Although it is not possible to isolate Effect~(1) in the simulation, events where the projectile does not interact with the target nucleus can be selected to isolate Effect~(2).
We analyze such non-interacting events and fit the time-evolved spatial distributions of the nucleons in the target nucleus to a Woods--Saxon form.
The resulting time-dependent Woods--Saxon parameters are shown in Fig.~\ref{fig:t_dep_of_ws}.
In PHSD, the simulation is performed in the center-of-mass frame of one nucleon from the projectile and one nucleon from the target.
Figure~\ref{fig:t_dep_of_ws} shows the dependence on the time in this center-of-mass frame.
The effects of Lorentz boost were taken into account in obtaining these parameters, making them comparable to those of a Woods--Saxon distribution in the laboratory frame.

Using both the time-dependent and static Woods-Saxon profiles, we repeated the analysis described in Sect.~\ref{sec:analysis} for the seven combinations of $(k_{1},k_{2})$ listed in Tables~\ref{tab:k1k2_common}~and~\ref{tab:k1k2_betagamma}.
The impact of the artificial deformation (Effect~(2)) was estimated from the difference in $\chi^{2}$ values obtained between the time-dependent and static cases.
These $\Delta\chi^{2}$ values were then used to estimate the corresponding systematic uncertainties in $k_{1}$ and $k_{2}$ due to target instability (see Table~\ref{tab:tgt_stability}).
\begin{table}[t]
\centering
\caption{
    Absolute differences in $\chi^{2}$ between the time-dependent and static Woods--Saxon cases.
    The corresponding $(k_{1},k_{2})$ values are those given in Tables~\ref{tab:k1k2_common}~and~\ref{tab:k1k2_betagamma}.
}
\label{tab:tgt_stability}
\begin{tabular}{ccccc}
\hline
Partial decay width & Slow & Middle & Fast & All $\beta\gamma$ \\
\hline
Density dependent & 0.55 & 0.06 & 0.37 & 0.86 \\
Constant          & -    & 0.11 & 0.30 & 0.25 \\
\hline
\end{tabular}
\end{table}

\subsubsection{Intrinsic $k_{2}$ of PHSD}
\noindent
As described in Sect.~\ref{sec:lifetime}, due to the complexity of the absorption processes within PHSD, the effective parameter $k_{2}^{\text{PHSD}}$, which characterizes the density dependence of the $\phi$ meson lifetime in the PHSD model, is not constant.
As shown in Fig.~\ref{fig:rho_phsd_k2}, the value of $k_{2}^{\text{PHSD}}$ varies significantly from 0.0 to approximately 10.0 depending on the local baryon density and the Lorentz factor.

To estimate the systematic uncertainty associated with the fixing of $k_2$ in the present calculation, we repeated the analysis described in Sect.~\ref{sec:analysis} using fixed values of $k_{2}^{\text{PHSD}}=0\text{ and }10$ as two extreme cases, in addition to the best-fit value of 7.18.
For each case, we searched for the $k_{1}$ and $k_{2}$ values that minimize $\chi^{2}$.
The resulting modification parameters are summarized in Table~\ref{tab:intrinsic_k2}.
\begin{table}[t]
\centering
\caption{
    Best-fit modification parameters for different fixed values of $k_{2}^{\text{PHSD}}$.
    The value of $k_{2}^{\text{PHSD}}=7.18\text{, }0\text{, and }10$ correspond to the best-fit value and the two extreme cases, respectively, as described in Sect.~\ref{sec:lifetime}.
}
\label{tab:intrinsic_k2}
\begin{tabular}{cccccccccc}
\hline
&& \multicolumn{2}{c}{Slow} & \multicolumn{2}{c}{Middle} & \multicolumn{2}{c}{Fast} & \multicolumn{2}{c}{All $\beta\gamma$} \\
$\Gamma^{ee}$ & $k_{2}^{\text{PHSD}}$ & $k_{1}$ & $k_{2}$ & $k_{1}$ & $k_{2}$ & $k_{1}$ & $k_{2}$ & $k_{1}$ & $k_{2}$ \\
\hline
$\rho$ dep. & 7.18 & 0.026 & 7.9 & --0.019 & 0.2 & 0.008 & 0.0 & 0.000 & 2.2 \\
            & 0    & 0.026 & 6.8 & --0.013 & 0.0 & 0.008 & 0.0 & --0.002 & 1.6 \\
            & 10   & 0.021 & 8.9 & --0.017 & 0.0 & 0.008 & 0.0 & --0.001 & 2.0 \\
\hline
Const. & 7.18 & - & - & --0.018 & 0.0 & 0.006 & 0.0 & 0.008 & 11.0 \\
       & 0    & - & - & --0.017 & 0.0 & 0.006 & 0.0 & 0.005 & 9.8 \\
       & 10   & - & - & --0.020 & 0.1 & 0.008 & 0.0 & 0.003 & 8.7 \\
\hline
\end{tabular}
\end{table}

\subsubsection{Background shape}
\noindent
To estimate the systematic uncertainty associated with the choice of the background function in the spectral fitting, we varied both the functional form and parameter determination method.
In the default analysis, an exponential background was used, and the parameters were determined by excluding the region \SI{0.95}{\giga\electronvolt/\textit{c}^{2}}--\SI{1.05}{\giga\electronvolt/\textit{c}^{2}}.
To assess the sensitivity of the fit to those assumptions, two alternative scenarios were considered:
\begin{enumerate}
    \item{The background shape was changed from exponential to a quadratic function.}
    \item{The excluded region used to determine the background parameters was widened to \SI{0.9}{\giga\electronvolt/\textit{c}^{2}}--\SI{1.1}{\giga\electronvolt/\textit{c}^{2}}.}
\end{enumerate}
For each case, the $k_{1}$ and $k_{2}$ parameters were re-optimized using the procedure described in Sect.~\ref{sec:analysis}.
The resulting values are summarized in Table~\ref{tab:bg_shape}.
\begin{table}[t]
\centering
\caption{
    Best-fit modification parameters under different background shape assumptions.
    In Case~(0), the background is determined using an exponential fit, excluding the \SI{0.95}{\giga\electronvolt/\textit{c}^{2}}--\SI{1.05}{\giga\electronvolt/\textit{c}^{2}} region, which is the procedure used to obtain the final value.
}
\label{tab:bg_shape}
\begin{tabular}{cccccccccc}
\hline
&& \multicolumn{2}{c}{Slow} & \multicolumn{2}{c}{Middle} & \multicolumn{2}{c}{Fast} & \multicolumn{2}{c}{All $\beta\gamma$} \\
$\Gamma^{ee}$ & Background & $k_{1}$ & $k_{2}$ & $k_{1}$ & $k_{2}$ & $k_{1}$ & $k_{2}$ & $k_{1}$ & $k_{2}$ \\
\hline
$\rho$ dep. & (0) & 0.026 & 7.9 & --0.019 & 0.2 & 0.008 & 0.0 & 0.000 & 2.2 \\
            & (1) & 0.029 & 4.7 & --0.017 & 0.0 & 0.003 & 0.0 & --0.007 & 0.0 \\
            & (2) & 0.017 & 11.7 & --0.021 & 0.0 & 0.007 & 0.0 & --0.001 & 3.4 \\
\hline
Const. & (0) & - & - & --0.018 & 0.0  & 0.006   & 0.0 & 0.008   & 11.0 \\
       & (1) & - & - & --0.020 & 0.0  & --0.001 & 0.0 & --0.006 & 0.0 \\
       & (2) & - & - & --0.024 & 10.0 & 0.004   & 0.0 & 0.011   & 17.1 \\
\hline
\end{tabular}
\end{table}

\subsection{Result with systematic uncertainty}
\label{sec:result_with_errors}
\noindent
The results of our analysis including the different sources of the systematic uncertainty are summarized in Table~\ref{tab:k1k2_w_error}.
\begin{table}[t]
\centering
\caption{
    Best-fit modification parameters for each $\beta\gamma$ region and partial decay width.
    The quoted errors are: statistical (first), PHSD-related systematic (second), and background-related systematic (third).
    Statistical uncertainties correspond to the upper and lower limits of the $\Delta\chi^{2}=1.00$ region.
    Asymmetric statistical uncertainties in $k_{1}$ arise due to the constraint $k_{2}\geq 0$.
    If $k_{2}=0$ lies within the statistical uncertainty, a 90\% confidence level upper limit is quoted, with the central value shown in parenthesis.
    The corresponding values of Ref.~\cite{muto}, which only include statistical errors, are also given for comparison.
}
\label{tab:k1k2_w_error}
\begin{tabular}{cccc}
\hline
Partial decay width & $\beta\gamma$ & $k_{1}$ & $k_{2}$ \\
\hline
Density dependent & Slow & $0.026^{+0.021+0.015+0.003}_{-0.021-0.015-0.008}$ & $7.9\pm3.9^{+2.9+3.8}_{-2.9-3.2}$ \\
& Middle & $-0.019^{+0.016+0.006+0.002}_{-0.014-0.004-0.002}$ & $\leq 4.8$ ($0.2^{+2.8+0.7+0.0}_{-0.2-0.2-0.2}$) \\
& Fast & $0.008^{+0.016+0.010+0.000}_{-0.017-0.010-0.006}$ & $\leq 2.6$ ($0.0^{+1.6+1.0+0.0}$) \\
& All $\beta\gamma$ & $0.000^{+0.010+0.009+0.000}_{-0.010-0.009-0.007}$ & $\leq 6.8$ ($2.2^{+2.8+2.6+1.2}_{-2.2-2.2-2.2}$) \\
& Ref.~\cite{muto} & $0.034^{+0.006}_{-0.007}$ & $2.6^{+1.8}_{-1.2}$ \\
\hline
Constant & Slow & - & - \\
& Middle & $-0.018^{+0.022+0.008+0.000}_{-0.023-0.008-0.004}$ & $\leq 32.2$ ($0.0^{+19.5+6.5+9.0}$) \\
& Fast & $0.006^{+0.017+0.009+0.000}_{-0.017-0.009-0.007}$ & $\leq 6.8$ ($0.0^{+4.1+2.3+0.0}$) \\
& All $\beta\gamma$ & $0.008^{+0.018+0.009+0.004}_{-0.018-0.009-0.013}$ & $11.0\pm8.7^{+4.3+6.8}_{-4.3-11.0}$ \\
& Ref.~\cite{muto} & $0.033^{+0.011}_{-0.008}$ & $0^{+5.6}$ \\
\hline
\end{tabular}
\end{table}
In the case of $\beta\gamma$-independent $k_{1}$ and $k_{2}$ (``All $\beta\gamma$" in Table~\ref{tab:k1k2_w_error}), the results assuming a constant partial width show non-zero width broadening beyond the estimated uncertainties.
When introducing $\beta\gamma$ dependence, both non-zero $k_{1}$ and $k_{2}$ exceeding statistical and systematic errors were observed in the slow and middle $\beta\gamma$ regions for the density-dependent partial width case.

To quantitatively compare model validity, we carried out a likelihood ratio test for the density-dependent partial width case.
The number of parameters and the sum of the corresponding $\chi^{2}$ values for $\beta\gamma$-independent and $\beta\gamma$-dependent model are 8 and 12, and 343 and 327, respectively.
Using these values, the $\beta\gamma$-independent model for the density-dependent case is rejected at the 99\% confidence level.
This conclusion is consistent with the improved agreement in Fig.~\ref{fig:each_param_fit_result} compared to Fig.~\ref{fig:common_param_fit_result}.

\subsection{Comparison with theoretical calculation and other experiment}
\begin{figure}[!h]
\centering
\includegraphics[width=170mm]{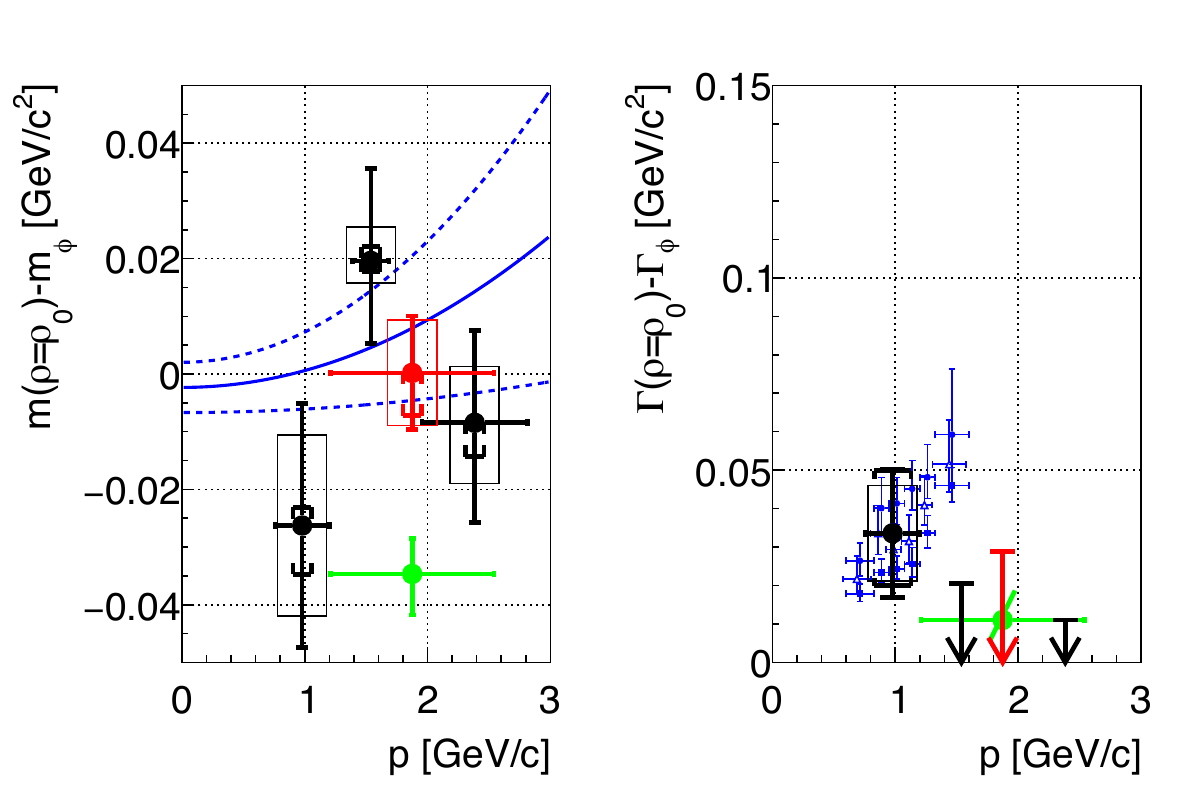}
\caption{
    Momentum dependence of $\phi$ meson resonance mass (left) and width (right) at normal nuclear density, for the case with the density-dependent partial width.
    Black (red) points indicate the results of the $\beta\gamma$-dependent (independent) model in the present analysis, while green points show the results from Ref.~\cite{muto}.
    Black and red vertical error bars shown as lines, squares, and brackets, represent statistical uncertainties, systematic uncertainties from the PHSD simulation, and systematic uncertainties due to background modeling, respectively.
    Horizontal error bars represent the RMS of the momentum distribution.
    Green points are calculated from the $k_{1}$ and $k_{2}$ values reported in Ref.~\cite{muto}.
    Green error bars represent statistical uncertainties only.
    Black and red arrows in the right panel indicate upper limits.
    Blue solid curve in the left panel represents theoretical predictions, with father details provided in the main text.
    Blue dashed curves in the left panel indicate the uncertainty of these theoretical predictions.
    Blue points with error bars in the right panel correspond to the COSY-ANKE results digitized from Fig.~4(a) of Ref.~\cite{cosy_anke}.
    The blue circles, squares, and open triangles represent different theoretical models employed in the analysis \cite{cosy_anke_model1,cosy_anke_model2,cosy_anke_model3}.
    To improve visual clarity, certain error bars are displayed at an angle.
}
\label{fig:mom_mass_width_rho_dep}
\end{figure}

\noindent
The momentum dependence of the $\phi$ meson resonance mass and width at normal nuclear density, for the case with the density-dependent partial width, as obtained from the present analysis, is shown in Fig.~\ref{fig:mom_mass_width_rho_dep}.
The green curves in Fig.~\ref{fig:mom_mass_width_rho_dep} are derived as follows, based on QCD sum rule and lattice QCD calculations.
The functional form
\begin{equation}
    m(\rho=\rho_{0},p)-m(\rho=0,p=0)=a+bp^{2}
\end{equation}
is based on the QCD sum rules, specifically Eq.~(50) of Ref.~\cite{Kim:2019ybi}.
The momentum-dependent coefficient $b$ was calculated from an appropriately weighted average of transverse and longitudinal modes, using $b=\frac{2b^{\text{T}}+b^{\text{L}}}{3}$, where $b^{\text{T}}$ and $b^{\text{L}}$ are the transverse and longitudinal components, respectively, as provided in Ref.~\cite{Kim:2019ybi}.
The coefficient $a$, corresponding to the mass difference between vacuum and normal nuclear density at zero momentum, was obtained using $a=b_{0}-b_{1}\sigma_{sN}$ (from Eq.~(13) of Ref.~\cite{sum_rule1}), with $\sigma_{sN}$ being the strange sigma term $\sigma_{sN} = m_s \langle N|\bar{s}s|N\rangle$ (which can alternatively be understood as the product of the strange quark mass and the leading-order coefficient in the small density expansion of the strange quark condensate).
The values of $b_{0}$ and $b_{1}$ were taken from Ref.~\cite{sum_rule1} and $\sigma_{sN}$ was adopted as the average of Eqs.~(449)~and~(450) in Ref.~\cite{flag2024}, which are averaged values of recent lattice QCD calculations of this quantity.
The results from Ref.~\cite{muto} fall outside the theoretical band, whereas the present results are closer to the theoretical predictions and lie within their uncertainty ranges.

The momentum dependence was also compared with a measurement of the $\phi\rightarrow K^{+}K^{-}$ cross section from the COSY-ANKE experiment \cite{cosy_anke}.
Since this experiment provides data only up to about \SI{1.5}{\giga\electronvolt/\textit{c}}, only the slow and middle $\beta\gamma$ regions are comparable. 
Comparing the findings of Ref.~\cite{cosy_anke} with ours, it is observed that the results in the middle $\beta\gamma$ region deviate beyond the uncertainties, while those in the slow $\beta\gamma$ region show good agreement with the COSY-ANKE data.

\begin{figure}[!h]
\centering
\includegraphics[width=170mm]{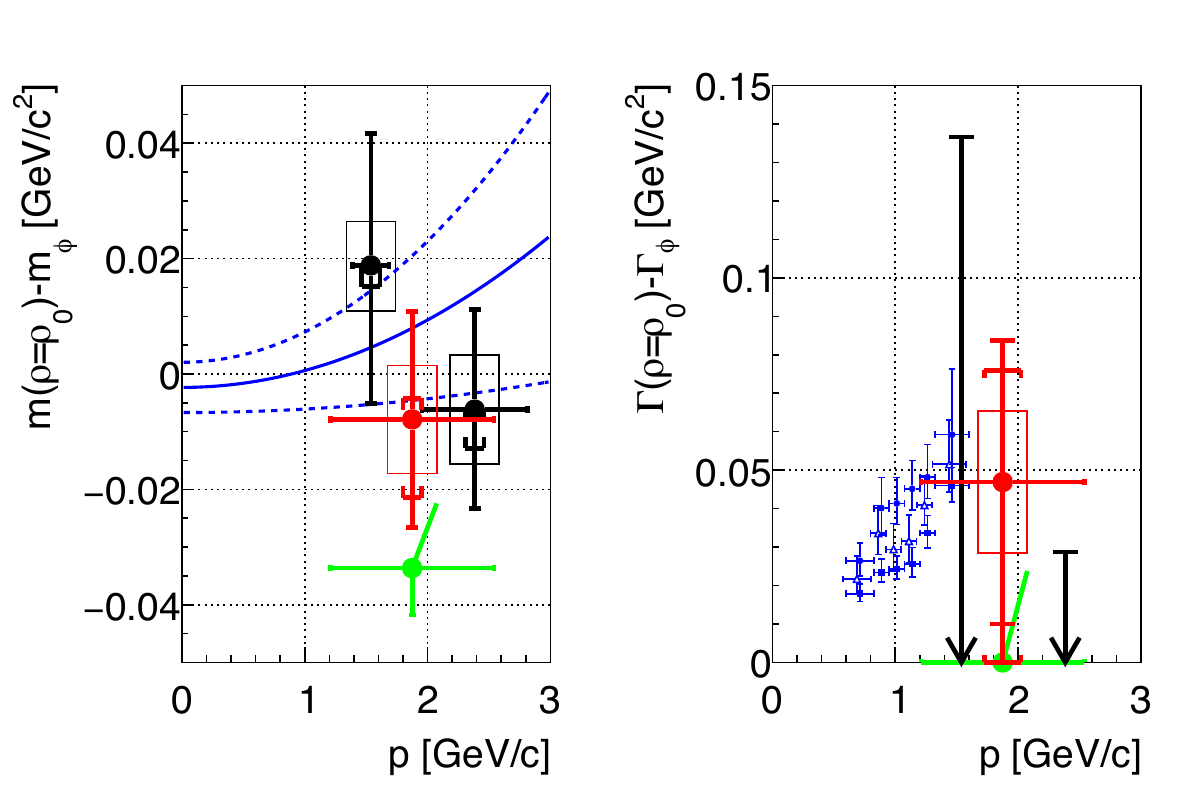}
\caption{
    Momentum dependence of $\phi$ meson resonance mass (left) and width (right) at normal nuclear density, for the case with the constant partial width.
    Points, lines, and arrows are defined in the same way as in Fig.~\ref{fig:mom_mass_width_rho_dep}.
}
\label{fig:mom_mass_width_const}
\end{figure}

Figure~\ref{fig:mom_mass_width_const} presents the results for the constant partial width case, where the extracted ($k_{1}$, $k_{2}$) values for slow $\phi$ mesons were found to be unphysical.

\subsection{Comparison with previous analysis}
\begin{figure}[!h]
\centering
\includegraphics[width=170mm]{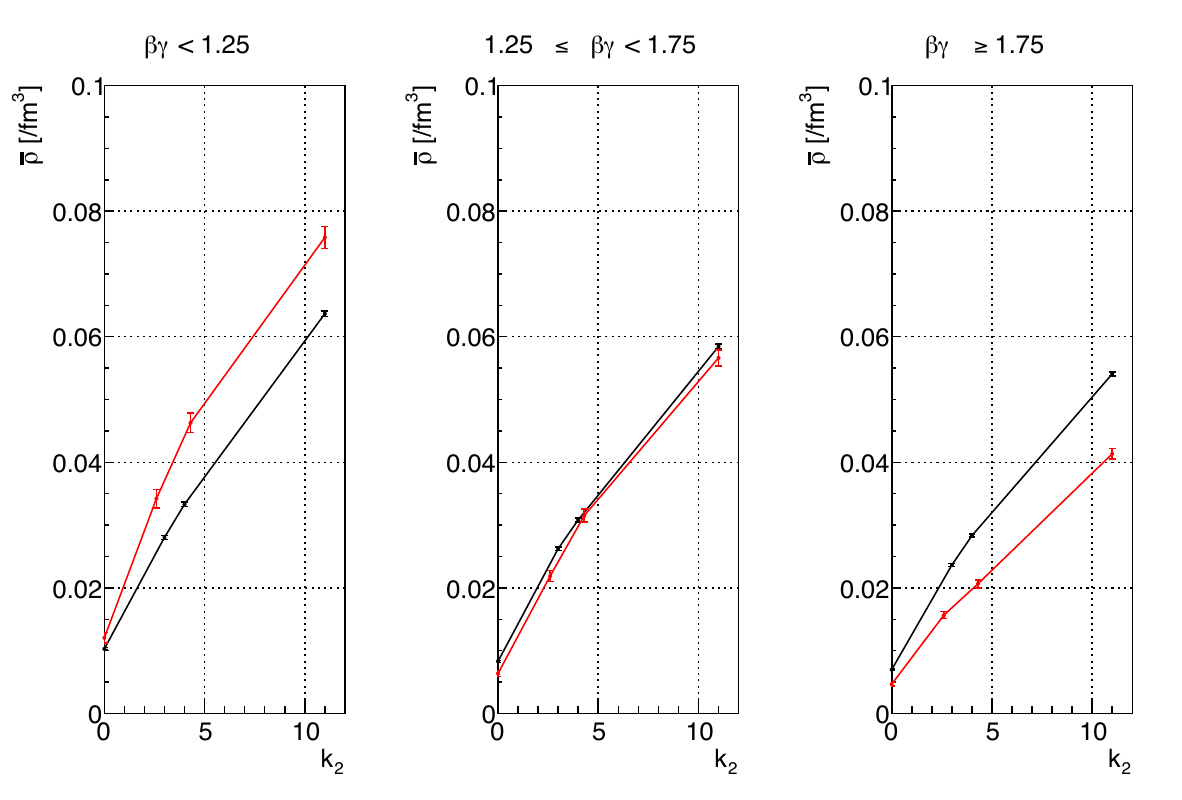}
\caption{
    Dependence of the average baryon density at the $\phi$ meson decay point on $k_{2}$ for the Cu target.
    The black points represent results from present simulation, while the red points show results from a simulation setup equivalent to Ref.~\cite{muto}.
}
\label{fig:k2_rho}
\end{figure}

\noindent
Compared with Ref.~\cite{muto} in the case of the $\beta\gamma$-independent model, only $k_{1}$ in the case of the density-dependent width is inconsistent within the uncertainties.
This discrepancy arises from the greater influence of the fast $\beta\gamma$ region compared to Ref.~\cite{muto}, leading to a reduced best-fit value for $k_{1}$.
To better understand this result, Fig.~\ref{fig:k2_rho} shows a comparison of the $k_{2}$ dependence of the average density at the $\phi$ meson decay point in each $\beta\gamma$ region with that of Ref.~\cite{muto}.
In the slow $\beta\gamma$ region, the average density of the present study is lower than in Ref.~\cite{muto}, while in the fast $\beta\gamma$ region, it is higher.
This increased density in the fast region enhances the effect of $k_{1}$ and $k_{2}$, but the corresponding experimental spectrum shows little to no modification.
As a result, both $k_{1}$ and $k_{2}$ in the $\beta\gamma$-independent ($k_{1}$, $k_{2}$) model are effectively suppressed in the present analysis.

In the $\beta\gamma$-independent model, the results of the previous analysis and the present one are inconsistent, as shown in Table~\ref{tab:k1k2_w_error}.
However, as discussed in Sect.~\ref{sec:result_with_errors}, the $\beta\gamma$-independent model is rejected in the case of the density-dependent partial width.
Therefore, we compared the values of $k_{1}$ and $k_{2}$ obtained solely from the Cu data in the slow $\beta\gamma$ region.
These results are consistent with the previous analysis \cite{muto_d}, as shown in Table~\ref{tab:cu_slow}.
\begin{table}[t]
\centering
\caption{
    Best-fit modification parameters for the Cu data in the slow $\beta\gamma$ region in the case of density-dependent partial width.
    Only statistical uncertainties are considered.
}
\label{tab:cu_slow}
\begin{tabular}{cccc}
\hline
& $k_{1}$ & $k_{2}$ & $\chi^{2}/\text{ndf}$ \\
\hline
present analysis & $0.034\pm0.012$ & $8.5\pm3.3$ & 74/51 \\
Ref.~\cite{muto_d} & $0.031^{+0.005}_{-0.003}$ & $6.1^{+2.3}_{-1.5}$ & 63.4/48 \\
\hline
\end{tabular}
\end{table}

\section{Conclusion}
\label{sec:conclusion}
\noindent
In the KEK-PS E325 experiment, 
the \phiee\ invariant mass spectrum was measured
using 12 \SI{12}{\giga\electronvolt} proton beam and nuclear targets of C and Cu. A significant spectral modification
was observed.
In the model calculation used in the previous analysis \cite{muto} to compare the data with theoretical
predictions on the spectral modification in dense matter, 
the time evolution of the spatial density distribution was not taken into account, and the production points of the $\phi$ mesons were assumed to be distributed
proportionally to the density of the target nucleus.

To newly incorporate these effects, the present study employed the PHSD transport approach.
A similar analysis to Ref.~\cite{muto} was performed.
In the $\beta\gamma$-dependent parameter model, that is, different $(k_{1},k_{2})$ values for each $\beta\gamma$ region, the data are reproduced better than in the $\beta\gamma$-independent model.
The $\beta\gamma$ dependence of $(k_{1},k_{2})$, which has been theoretically suggested, is strongly supported by the present analysis.
Compared to theoretical calculation based on QCD sum rule and lattice QCD, the present results are within their uncertainty ranges, as shown in Fig.~\ref{fig:mom_mass_width_rho_dep}.

As shown in Table~\ref{tab:cu_slow}, the obtained modification parameters are consistent with the previous analysis for
the slow Cu data \cite{muto_d}, where the significant spectral modification is observed.

The uncertainties in the current analysis are primarily due to statistical limitations of the experiment, while the systematic uncertainties are subleading in most cases. 
Our results provided strong motivation for the high-statistics measurement at the J-PARC E16 experiment \cite{e16}, combined with updated transport simulations for those conditions.
This will reduce the still rather large uncertainties and will further clarify the finite-density behavior of the $\phi$ meson as a function of momentum.

\section*{Acknowledgment}

\noindent
We sincerely thank all the staff members of KEK, including the PS accelerator division, online group, electronics group, PS floor group and particularly the PS beam channel group for their helpful support.
We also express our gratitude to the staff members of RIKEN-CCJ, RIKEN RSCC,  RIKEN Hokusai, and KEKCC for their support on the data analysis, data archiving, and simulation.
In relation to the PHSD transport simulations, we express our gratitude to Profs.~E.~Bratkovskaya and  T.~Song, for their kind support and valuable discussion.
English proofreading was performed using ChatGPT.
This study was partly supported by the JSPS Research Fellowship for Young Scientists, RIKEN Special Postdoctoral Researchers Program, and MEXT/JSPS KAKENHI Grant Numbers JP06640391, JP07640396, JP08404013, JP12440064, JP15340089, JP20H05647 and JP23H05440.


%

\vspace{0.2cm}
\noindent


\let\doi\relax


\end{document}